\documentclass[sigconf, balance=false]{acmart}
\AtBeginDocument{%
  }
\acmISBN{978-1-4503-XXXX-X/2018/06}

\usepackage{graphicx}
\usepackage{tikz}
\usepackage{amsmath}
\usepackage{pifont}
\usepackage{url}
\usepackage{hyperref}
\usepackage{xspace}
\usepackage{mathbbol}
\usepackage{dsfont}
\usepackage{subcaption}
\usepackage{caption}
\usepackage{listings}
\usepackage[framemethod=TikZ]{mdframed}
\usepackage{multirow}
\usepackage{booktabs}
\usepackage{eso-pic}
\usepackage{tcolorbox}
\usepackage{color}
\usepackage{xcolor}

\long\def\comment#1{}

\usepackage{bbm}
\usepackage{enumitem}
\usepackage{colortbl}
\usepackage{float}
\hypersetup{
	colorlinks=true,
	linkcolor=blue,
	filecolor=blue,      
	urlcolor=black,
	citecolor=blue,
}

\begin{document}


\title[SmartGuard: Leveraging LLMs for Network Attack Detection through Audit Log Analysis and Summarization]{SmartGuard: Leveraging Large Language Models for Network Attack Detection through Audit Log Analysis and Summarization}
\renewcommand\footnotemark{}
\renewcommand\footnoterule{}

\author{Hao Zhang}
\affiliation{%
  \institution{Zhejiang University}
  \city{Hangzhou}
  \country{China}
}
\email{haozhang\_hz@zju.edu.cn}

\author{Shuo Shao}
\affiliation{%
  \institution{Zhejiang University}
  \city{Hangzhou}
  \country{China}
}
\email{shaoshuo\_ss@zju.edu.cn}

\author{Song Li}
\affiliation{%
  \institution{Zhejiang University}
  \city{Hangzhou}
  \country{China}
}
\email{songl@zju.edu.cn}

\author{Zhenyu Zhong}
\affiliation{%
  \institution{Ant Group}
  \city{Hangzhou}
  \country{China}
}
\email{edward.zhong@antgroup.cn}

\author{Yan Liu}
\affiliation{%
  \institution{Ant Group}
  \city{Hangzhou}
  \country{China}
}
\email{bencao.ly@antgroup.cn}

\author{Zhan Qin}
\affiliation{%
  \institution{Zhejiang University}
  \city{Hangzhou}
  \country{China}
}
\email{qinzhan@zju.edu.cn}


\begin{abstract}
\label{sec:abstract}
End-point monitoring solutions are widely deployed in today's enterprise environments to support advanced attack detection and investigation. These monitors continuously record system-level activities as audit logs and provide deep visibility into security events. Unfortunately, existing methods of semantic analysis based on audit logs have low granularity, only reaching the system call level, making it difficult to effectively classify highly covert behaviors. Additionally, existing works mainly match audit log streams with rule knowledge bases describing behaviors, which heavily rely on expertise and lack the ability to detect unknown attacks and provide interpretive descriptions. In this paper, we propose SmartGuard, an automated method that combines abstracted behaviors from audit event semantics with large language models. SmartGuard extracts specific behaviors (function level) from incoming system logs and constructs a knowledge graph, divides events by threads, and combines event summaries with graph embeddings to achieve information diagnosis and provide explanatory narratives through large language models. Our evaluation shows that SmartGuard achieves an average F1 score of 96\% in assessing malicious behaviors and demonstrates good scalability across multiple models and unknown attacks. It also possesses excellent fine-tuning capabilities, allowing experts to assist in timely system updates.


\end{abstract}



\maketitle

\section{Introduction}\label{sec:introduction}

Enterprise security today depends heavily on endpoint monitoring tools like SIEM \cite{logrhythm2024}, which track system-level execution via audit logs. While these tools are essential for preventing incidents like the Capital One and Twitter attacks \cite{capitalone2019, guardian2020}, they generate an overwhelming amount of information. Processing millions of log events daily \cite{gao2018saql} remains a major challenge, as the sheer scale of data often hides critical evidence of unauthorized access.
Although analysts rely on these logs for causal investigation \cite{king2003backtracking}, existing methods focus on log reduction and query optimization without interpreting semantic meanings \cite{liu2018timely}. This leaves a critical semantic gap between low-level events and high-level behaviors. Rule-based methods \cite{milajerdi2019poirot} attempt to bridge this gap but require substantial expert knowledge and lack generalizability across diverse, large-scale audit logs.

Recently, the capabilities of large language models (LLMs) in processing complex tasks offer a promising avenue for enhancing high-level behavior analysis in audit logs \cite{logrhythm2024, hossain2017sleuth}. These models can efficiently parse voluminous data, identify relevant patterns, and generate concise interpretations, thereby significantly reducing the manual effort required from analysts and accelerating incident response. Moreover, LLMs can adapt to novel threats by learning from historical data. For instance, models like OPT demonstrate cost-effective performance suitable for large-scale inference \cite{zhang2022opt}. They can automatically abstract and label semantically similar behaviors without pre-existing labels, allowing analysts to focus only on representative cases. Beyond reducing manual workload, LLMs also facilitate proactive detection of anomalous behaviors. However, applying LLMs to audit log analysis faces challenges such as accurately differentiating fine-grained event semantics and delineating behavior boundaries within highly interleaved event streams. Their lack of innate domain-specific knowledge further limits precision in root-cause analysis and behavior interpretation.

In this paper, we propose SmartGuard, an automated behavior abstraction and detection method that bridges the semantic gap in audit events by leveraging large language models. Unlike previous approaches that rely on expert knowledge, SmartGuard automatically derives contextual semantics directly from audit logs by preprocessing data, extracting behavioral subjects/objects, and categorizing logs by threads. It generates natural language summaries and uses Graph Neural Networks to create behavioral embeddings, aggregating event semantics into high-level behavioral representations. After fine-tuning a large language model on these abstractions, SmartGuard achieves accurate behavior diagnosis with interpretable narratives. Evaluation on the DARPA TC dataset demonstrates its effectiveness, with an average F1 score of 95.2\% in behavior abstraction and good scalability across models.

Our contributions are summarized as follows.
\begin{itemize}[leftmargin=*]
    \item We propose SmartGuard, a method that combines large language models with the extraction of behaviors from logs, enabling it to effectively detect both known and unknown attacks. Our approach summarizes behaviors guided by information flow and enables large language models to perform information diagnostics by aggregating contextual semantics.
    \item We propose a new scheme for abstracting contextual semantics, achieving fine granularity down to specific behaviors (function level), and dividing events by threads. This allows integration with large language models to provide explanatory narratives for behavior semantics.
    
    \item We conducted a systematic evaluation and scalability analysis of common malicious behaviors. The results show that SmartGuard can effectively abstract high-level behaviors and detect them, demonstrating good scalability across multiple models and unknown attacks. Additionally, it can efficiently collaborate with experts for fine-tuning and real-time system updates.
\end{itemize}

\section{Background and Motivation}\label{sec:background}


\subsection{Motivating Example}

\begin{figure}[t]
    \centering
    \includegraphics[width=\linewidth]{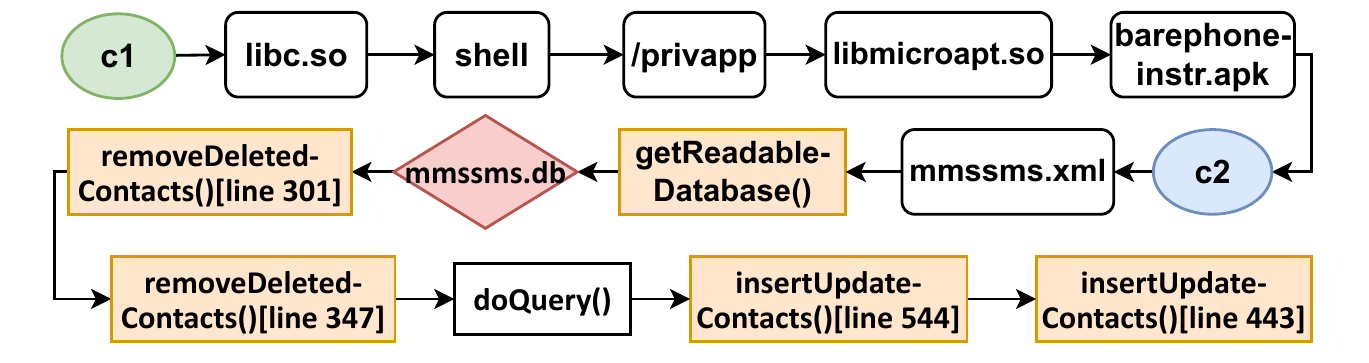}
    \caption{Scenario example. The nodes in the figure are system entities (rectangles represent functions, rounded rectangles represent addresses and files, ellipses represent sockets, and diamonds represent databases). The edges between the nodes represent system calls. For clarity, we color-code the source data objects in the behavior, with red and yellow representing high-risk behaviors.}
    \label{fig:example1}
    \vspace{-0.3cm}
\end{figure}

Consider the attack scenarios of Barephone Micro. As an attacker, you write a malicious application APK intending to steal database files. The victim user installs and runs the malicious APK, which loads the Micro APT shared object. Micro APT connects to \texttt{77.138.117.150:80} as \texttt{C2}. The address \texttt{128.55.12.114} is used as \texttt{C1}, where a benign activity installs an elevation driver and uses the driver for privilege escalation. Finally, the new permissions were used to call \texttt{getReadableDatabase} to steal the database files \texttt{mmssms.db}. Multiple calls to \texttt{removeDeletedContacts} were then made to clear the relevant information from the database.
Although this strategy executes high-risk functions multiple times, traditional schemes do not deliberately record specific function behaviors, leading analysts to mistakenly classify it as a normal event. Figure \ref{fig:example1} shows the behavior diagram of the strategy.

\subsection{Challenges}


Security analysts must identify both malicious actions (e.g., data breaches) and benign activities (e.g., file uploads) when investigating attacks. While provenance graphs help visualize event dependencies and filter irrelevant data, analysts still waste time examining benign events from common activities. Abstracting behaviors from audit events allows analysts to focus on meaningful patterns instead of individual events. This approach reduces the analysis scope from the full event set to a manageable number of salient behaviors. However, to automatically abstract high-level behaviors from low-level audit events and accurately classify them, analysts face three major challenges:

\begin{itemize}[leftmargin=*]
    \item \textit{Refining the semantics of audit events.} Audit records are usually based on processes as the smallest unit, but they overlook the specific functions executed in each step of behavior. This can make certain malicious behaviors difficult to detect. For example, the theft of files from two entities with different names but similar semantics may not be recognized.
    \item \textit{Inferring the semantics of audit events and identifying behavior boundaries.}  Audit events capture low-level system execution states but lack high-level semantic context, making behavior recognition challenging. For instance, identical system entity names may imply different intents. Current methods often rely on expert-defined rules or knowledge bases to infer event semantics, but manual approaches hinder scalability given the volume and interleaved nature of audit data—e.g., a single APT installation can generate over 30,000 events. Furthermore, dense causal relationships across events complicate the segmentation of individual behaviors and the identification of behavior boundaries.
    \item \textit{Identifying unknown attacks and achieving timely system updates. } Although existing solutions can recognize known attack behaviors, unknown attacks occur more frequently in reality. Unlike known attack types, learning the semantic relationships between behaviors and recognizing unknown attacks is a challenge. More importantly, real-world systems should be capable of interacting with security personnel and promptly updating the types of attacks they recognize.
\end{itemize}

\subsection{Problem Analysis}
Given a large number of audit logs in a user login session, our goal is to identify high-level (both benign and malicious) behaviors and provide a quantitative representation of their semantics without analyst involvement. Additionally, we aim to use LLMs to identify and provide interpretable narratives for abstract high-level behaviors. Compared to traditional methods of behavior abstraction that heavily rely on domain knowledge, our goal is to achieve automated behavior abstraction using simple and effective insights.

\begin{figure}[t]
    \centering
    \includegraphics[width=\linewidth]{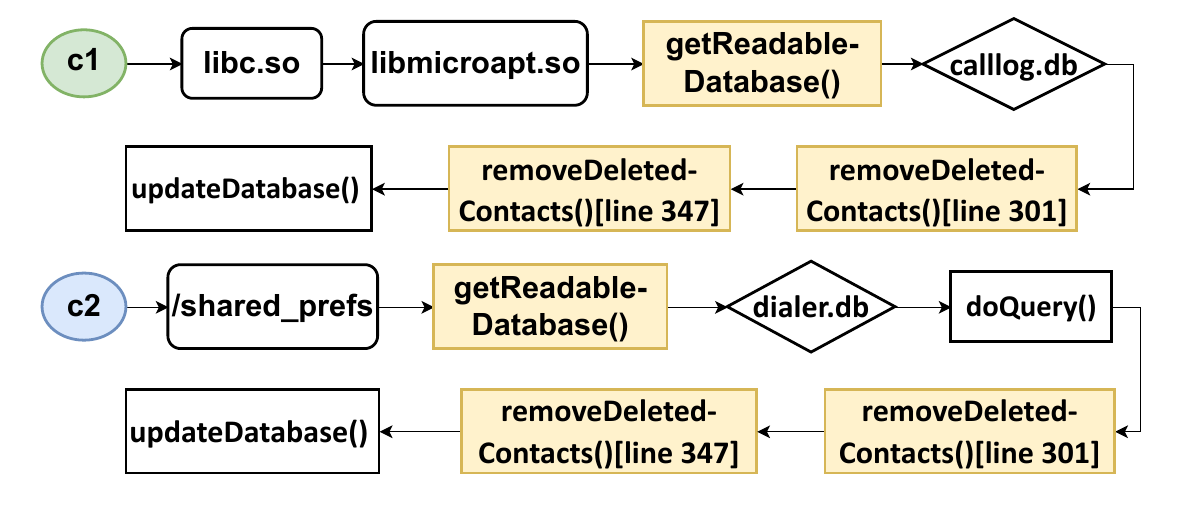}
    \caption{Attack subgraph for stealing information from different databases. We color-coded the data objects, with yellow indicating similar behavior semantics.}
    \label{fig:example2}
    \vspace{-0.5cm}
\end{figure}

Our first insight is that function-level behavior recording and thread-based log division can reveal more covert behaviors. Traditional audit logs represent events as (Subject, Relation, Object) triplets based on processes, which capture high-level actions but omit fine-grained details such as function parameters and specific object attributes. These details, often embedded in Object tags, are critical for detecting malicious behaviors—for instance, when an attacker invokes specific functions like \texttt{getReadableDatabase()} to access sensitive files such as \texttt{calllog.db}. Such subtle attacks remain hard to detect under process-based abstraction. Moreover, since related actions often occur within the same thread, thread-based segmentation offers a more meaningful unit for behavior analysis than process-oriented division. Figure \ref{fig:example2} illustrates examples of such behaviors.

Our second insight is that the semantics of system entities and relations in audit events can be revealed from their usage context. Similarly, for the other files, such as \texttt{calllog.db}, the read operation will call the \texttt{getReadableDatabase()} function. Even though different behaviors include other nodes, we can still determine their similarity from the overall context. Therefore, we can infer that despite different identifiers, these files may share similar semantics.

The core idea is to reveal the semantics of system entities and relations from the contextual information in audit events, such as by analyzing their correlations in events. A general method for extracting such contextual semantics is to use graph embedding models. It is to map system entities and relations into a graph embedding space (i.e., numerical vector space), where the distances between vectors capture semantic relationships. At the same time, generating log summaries from segmented events and integrating them with graph vectors becomes the event semantics. Now we can interpret the semantic information of audit events. The next step is to identify audit events that belong to malicious behaviors.

\begin{figure*}[ht]
    \centering
    \includegraphics[width=0.95\textwidth]{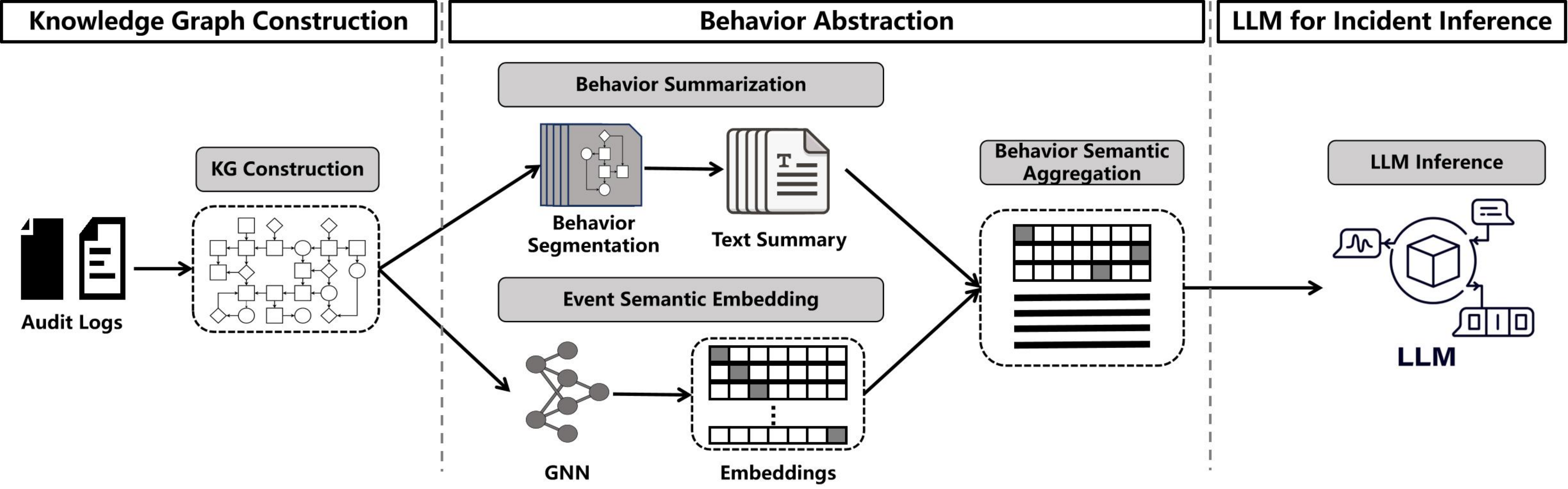} 
    \caption{SmartGuard Overview. First, we extract specific behaviors (function-level) from the logs and construct a knowledge graph. Second, we divide the behaviors according to threads and extract text summaries. Then, we perform embeddings on the extracted behavior subgraphs and combine them with text to form behavior semantics. Finally, we use a large language model to diagnose the behavior semantics and provide explanatory narratives.}
    \label{fig:overview}
    \vspace{-1em}
\end{figure*}

Our third insight is that large language models can be used to identify high-level log semantics. Based on our observations, large language models have achieved remarkable success in inference tasks. Although large language models have limited knowledge in the area of audit logs, their superior performance in fine-tuning allows for improved semantic recognition through the \textit{Chain-Of-Thought (COT)} approach. For example, by asking guiding questions such as `What are the sensitive links in the log behavior?', `What are the sensitive nodes in the log behavior?', and `What is the malicious type of the log behavior?', we can enhance the accuracy and scalability of large language models in recognizing log semantics.

\subsection{Threat Model}

We assume that the underlying operating system, audit engine, and monitoring data are part of the Trusted Computing Base (TCB). Ensuring the integrity of the operating system kernel, endpoint monitors, or the audit logs themselves is beyond the scope of this work. This threat model is shared in related work on system auditing \cite{gehani2012spade}, \cite{hassan2019nodoze}, \cite{hossain2017sleuth}, \cite{hossain2020combating}, \cite{liu2018towards}, \cite{milajerdi2019holmes}. We also assume that behaviors are audited at the kernel level and their operations are captured as system call audit logs. Although attackers may attempt to perform malicious actions without executing any system calls to hide their tracks, such behaviors appear to be rare and their impact on other parts of the system is limited \cite{wagner2002mimicry}. In this paper, we focus on behaviors within single-user sessions. Our insights generally apply to cross-session behaviors.


\section{Methodology}\label{sec:overview}


\subsection{Approach Overview}\label{subsec:3.1}
The overall approach of SmartGuard is depicted in Figure \ref{fig:overview}. It consists of three main stages: Knowledge Graph Construction, Behavior Abstraction, and LLM for Incident Inference. SmartGuard takes system audit data as input, such as Linux audit logs \cite{linux_audit_kernel}. It summarizes behavior instances, abstracts their high-level semantics, and ultimately outputs diagnostic results and explanatory narratives.

Specifically, taking audit logs from user sessions as input, the Knowledge Graph Construction module first parses the logs into triples, divides them by threads, and constructs a log-based Knowledge Graph (KG). Then, the Event Semantic Inference module employs a graph neural model to infer the contextual semantics of nodes in the knowledge graph. Meanwhile, the Behavior Summarization module enumerates subgraphs from the knowledge graph to generate textual summaries of the logs. Combining node semantics, the Behavior Semantic Aggregation module subsequently enhances the subgraphs to encode the semantics of behavior instances. Finally, the Large Language Model (LLM) module performs information diagnosis based on the aggregated semantics and provides explanatory narratives, which can further reduce the workload of downstream tasks. We will introduce the design details of SmartGuard in the following sections.

\subsection{Behavior Abstraction}\label{subsec:3.3}

\subsubsection{Behavior Summarization}\label{subsec:3.3.1}
The first step in behavior abstraction is to identify behavior semantics from user login sessions, where a behavior instance consists of coherent audit events operating on related data. Existing approaches often analyze individual log events or their elements \cite{du2017deeplog, shen2018tiresias, shen2019attack2vec}, but such fine granularity only captures one-step actions, failing to retain semantic context across longer behavioral trajectories. We observe that behavioral themes are thread-local, and threads—typically containing fewer than ten events—provide a more meaningful granularity for semantic analysis. By aggregating events within a thread, we form thread-level events and reduce the problem of identifying a behavior instance to extracting causally connected subgraphs rooted at thread objects in the session knowledge graph. We also account for multi-branch data transmission behaviors where single paths may be insufficient.

First, we identified a list of sensitive key nodes by analyzing the frequency of node occurrences in the DARPA ground truth dataset.(To reduce false positives, common benign nodes such as "login" were excluded.) Additionally, within each defined thread, we calculated the in-degree and out-degree of nodes to determine key nodes specific to that thread. During the subsequent event merging process, both types of key nodes were taken into consideration.

Eigenvector centrality is a method for measuring the importance of nodes in a graph. It considers the importance of a node's neighbors and calculates the centrality value of each node through an iterative method. The formula for eigenvector centrality is:
\[C_E(v) = \frac{1}{\lambda} \sum_{u \in \mathcal{N}(v)} A_{vu} C_E(u). \tag{1}
\]
\( C_E(v) \) represents the eigenvector centrality of node \( v \).
    \( \lambda \) is the eigenvalue.
     \( \mathcal{N}(v) \) is the set of neighbors of node \( v \).
    \( A_{vu} \) is the element in the adjacency matrix \( A \) that represents the connection between node \( v \) and node \( u \). If there is an edge between node \( v \) and node \( u \), then \( A_{vu} = 1 \); otherwise, \( A_{vu} = 0 \).
     \( C_E(u) \) represents the eigenvector centrality of node \( u \).

For each thread, we first construct the adjacency matrix \( A \) of the graph, then calculate the eigenvalues \( \lambda \) and eigenvectors \( \mathbf{v} \) of the adjacency matrix, and finally select the eigenvector corresponding to the largest eigenvalue as the key node.
Next, we process each thread. For each thread, a search is performed from its key nodes to explore all nodes within a $k$ range. If this search encounters anothor key node belonging to a different thread(which are defined as nodes with high in/out-degrees or those on a sensitive node list).Then the two threads are deemed to have highly correlated behaviors. Consequently, they are merged into a single event.



Additionally, we also observe that malicious behavior generally ends within a few steps. Therefore, we believe that only when the radius \( k \) of a key node contains other key nodes, the behaviors of the two threads are deeply related; otherwise, the behavior in the other thread does not have a significant impact on the behavior of this thread.

Afterward, we first divide the behavior according to the threads, then expand the search to form new behaviors. If one behavior is a subset of another, the two behaviors will be further merged. Then, we generate corresponding natural language summaries based on the behaviors. Considering that we will eventually use large language models to achieve information diagnosis, and the graph embedding vectors do not contain text vector content recognizable by language models, we need to provide text information to the large language model. Thanks to the thread division scheme adopted by SmartGuard, there is not much node information within each division (generally no more than ten). Large language models such as GPT and Copilot can be used to convert graph information into natural language summaries. For example, \textit{(path1, read, getReadableDatabase(), write, b.txt)} can be organized as \textit{`path1 uses getReadableDatabase() to achieve read, writing to b.txt'}.

In summary, we apply a method based on key nodes and thread division to divide the session's KG into subgraphs, where each subgraph describes a behavior instance and generates a natural language summary for each subgraph.

\subsubsection{Event Semantic Embedding}\label{subsec:3.3.2}
Understanding the semantics of audit events is essential for abstracting high-level behaviors. We derive subgraph semantics from partitioned behavior instances, selecting this granularity as the basic unit for semantic reasoning to balance scalability and accuracy. To handle large-scale logs efficiently, we employ a computationally efficient embedding model that preserves semantic accuracy.

Inspired by word embedding techniques in NLP \cite{angeli2015leveraging, sutskever2014sequence} and their applications in system contexts \cite{chua2017neural, zeng2021watson}, we map thread-based event contexts into a vector space. This allows events with similar contexts—even those involving different system entities—to be represented proximally, capturing shared semantics. To achieve this, we use a Graph Attention Network (GAT) model to learn the mapping from thread events to semantic embeddings, ensuring that behaviorally similar events lie close in the vector space.

During embedding, we used the word2vec model. We used the Gensim \cite{rehurek2010software} library to generate node embeddings for training the graph. Gensim is a Python library for natural language processing, particularly adept at handling large-scale text data. Since each node in the source graph contains two attributes, type and name, the input feature \( f_u \) of the node \( u \) can be calculated as follows:
\[
f_u = \text{Concat}(w_u^{\text{type}}, w_u^{\text{name}}),  \tag{2}
\]
where \( w_u^{\text{type}} \) is the vector representation of the word for the type of node \( u \), such as a function or address, and \( w_u^{\text{name}} \) is the vector representation of the word for the name of node \( u \). 

The result of word2vec is an embedding vector that expresses the features of each node. We store it as the input to the graph neural network.

Next, we train the GAT model. Considering the difficulties brought by overly complex models, we prefer to implement our model with a simple network structure. We use contrastive learning in unsupervised learning to train the GAT model. During training, GAT minimizes the distance to the nearest node of the same type found in the KG, while maximizing the distance to randomly sampled nodes in the KG that are not connected and not of the same type. More Training details can be seen in Appendix \ref{app:lossfunction}



In summary, the result of GAT is an \(n \times m\) embedding matrix that maps \(n\) nodes into an \(m\)-dimensional embedding space. In our example, \(m\) is 128. Afterward, through the pooling model, we can obtain the graph embedding of the entire thread event.

\subsubsection{Behavior Semantic Aggregation}\label{subsec:3.3.3}
After summarizing the behavior instances, we proceed to extract the semantics of the behavior instances. Recall that each behavior instance consists of audit events, whose semantics have been represented using high-dimensional vectors in an embedding matrix. We then naturally derive the semantics of the behavior instances by combining the partitions of the behavior instances. Additionally, natural language summaries of the behavior instances are generated to provide textual information for the semantics. To obtain the semantic representation of the final combined behavior instances, a simple approach is to aggregate the text summaries of the constituent events with the embedding vectors. However, this method is not suitable when the text summaries are too long, as it would cause the summaries and vectors to have a biased influence on the final judgment of the large language model, whereas they should actually contribute equally. Therefore, to further simplify the behavior summaries, we integrate the noise events in the behavior.

\vspace{0.3em}
\noindent\textbf{Noise events.} 
Trivial events, such as routine file operations for edit history caching (\textit{vim, write, .viminfo}) or shell configuration reading (\textit{bash, read, /etc/profile}), are considered noise due to their periodic and system-routine nature rather than specific user behaviors. These events typically occur predictably and in a fixed order. To filter them, we identify all possible operations of a program, observe that most trivial events involve meaningless file reads/writes, and aggregate them under Read-List and Write-List nodes within the behavior graph. Unlike NODEMERGE \cite{tang2018nodemerge}, which focuses only on read operations during process initialization, our approach targets trivial events across all file operation types in general executions.

After denoising the behavior instances, the same noise behaviors will no longer appear in the event behavior summaries, significantly shortening the length of the text summaries without affecting the expression of key information. In the embeddings of the behaviors, although there are Read-List and Write-List nodes that can express certain trivial event semantics, they do not affect the importance of the overall behavior embeddings. 

Next, we aggregate the behavior text summaries with the embedding vectors. We construct the following sentence pattern: `The following is a summary of the log: [\textit{behavior text summary}], where the key point is [\textit{Key Node}], followed by the embedding of the behavior graph \textless\textit{Graph Embedding}\textgreater.' By simply aggregating and adding prompt language, we finally generate a log behavior description that the LLM can understand. Figure \ref{fig:prompt_input} shows the input format of the LLM during behavior inference.

In summary, the behavior abstraction phase uses a log-based knowledge graph as input to generate corresponding text summaries and graph embedding vectors, which are ultimately aggregated into descriptive information that the LLM can reason with.

\subsection{LLM for Incident Inference}\label{subsec:3.4}

Recent advances in LLMs have shown strong capabilities in contextual understanding and information generation. However, they often struggle with domain-specific or long-tail tasks such as root-cause analysis without explicit guidance \cite{chalkidis2023chatgpt, kasai2023evaluating}. To address this, SmartGuard leverages Chain-of-Thought (CoT) prompting—a gradient-free method that uses manual demonstrations with reasoning chains to steer the model toward logical inference.

Inspired by automatic prompt construction techniques \cite{zhang2023automatic}, we treat behavior summaries and root-cause labels as questions and reasoning steps. This allows the model to automatically identify sensitive node links within behavior graphs. We integrate both graph embeddings and textual behavior summaries into the CoT prompts, explicitly instructing the LLM to judge the malignancy of events and identify critical causal links while providing explanatory justifications, as shown in Figure \ref{fig:prompt_input}.


\begin{figure}[t]
  \centering
   \begin{tcolorbox}[colback=gray!30, colframe=gray!30, coltitle=black, arc=4pt, boxrule=0.5pt, boxsep=2pt, left=2pt, right=2pt, top=2pt, bottom=2pt, before skip=0.5\baselineskip, after skip=0.5\baselineskip]
 \textbf{Instruction:} The following description shows the log information of the event. It contains a summary of the log text and behavior Embeddings. Please focus on the Embeddings to determine whether this event is a malicious event. If it is, please provide the classification of the malicious event. 
 
\textbf{Input:} The following is a summary of the log: [\textit{Behavior Text Summary}], where the key nodes are [\textit{Key Node}], followed by the embedding of the behavior graph [\textit{Graph Embedding}]. 

\textbf{Output:} No. / Yes, the category is [\textit{Attack Type}].

 \end{tcolorbox}
  \caption{The prompt to predict incident category.}
  \label{fig:prompt_input}
  \vspace{-1em}
\end{figure}

In summary, the reasoning stage of large language models diagnoses abstract behavioral information and provides explanations through the CoT concept.




\section{Evaluation}\label{sec:evaluation}

\subsection{Implementation}\label{sec:implementation}

SmartGuard is implemented in Python 3.8 with approximately 2.5K lines of code. It ingests audit logs (e.g., Linux Audit formats) through a general input interface. The core of our approach involves constructing a thread-based Knowledge Graph (KG) where system events are converted into triplets (Head, Relation, Tail) and stored using Neo4j. We evaluate the system using the public DARPA TC dataset(in Appendix \ref{app:attackCase}), which contains APT attack scenarios and benign activities from a red team engagement. Further implementation details regarding embedding generation, hardware configuration, and dataset preprocessing are provided in Appendix \ref{app:implementationDetail}.

\begin{table*}[ht]
\tabcolsep=7mm
\renewcommand{\arraystretch}{1.05}
\centering
\caption{Evaluation results of detecting behavior abstraction in sessions. The \textbf{bold part} indicates the best detection metric F1-score for each category.
}
\label{tab:acc}
\scalebox{0.85}{
\begin{tabular}{c|ccc|ccc|ccc}
\toprule[1.5pt]
\multirow{2}{*}{Category}  & \multicolumn{3}{c|}{Extractor\cite{satvat2021extractor}(\%)}& \multicolumn{3}{c|}{SmartGuard w/ OPT-1.3b(\%)} & \multicolumn{3}{c}{SmartGuard w/ LLaMa2-3b(\%)} \\
\cline{2-10}
 & P & R  & F1 & P & R  & F1  & P & R & F1 \\
\hline
JAVA APK  & 91.0 & 97.0 & 93.9 & 94.7 & 93.9 & 94.3 & 95.6 & 94.5 & \textbf{95.0} \\
\hline
Barephone  & 88.0 & 100 & 93.6 & 95.9 & 94.5 & 95.2 & 97.2 & 96.8 & \textbf{97.0} \\
\hline
CADETS Nginx  & 100 & 84.0 & 91.3 & 100 & 100 & \textbf{100} & 100 & 98.9 & 99.4 \\
\hline
Firefox Drakon  & 100 & 90.0 & 94.7 & 100 & 95.6 & \textbf{97.8} & 100 & 95.2 & 97.5 \\
\hline
Metasploit  & 91.0 & 91.0 & 91.0 & 87.6 & 94.9 & 91.1 & 90.2 & 94.6 & \textbf{92.3} \\
\hline
Micro BinFmt  & 88.0 & 100 & 93.6 & 100 & 92.3 & \textbf{96.0} & 100 & 88.2 & 93.7 \\
\hline
AppStarter  & 100 & 88.0 & 93.6 & 97.8 & 96.1 & \textbf{96.9} & 94.2 & 97.4 & 96.8 \\
\hline
Webshell  & 89.0 & 89.0 & 89.0 & 92.4 & 93.3 & 92.8 & 95.4 & 98.2 & \textbf{97.1} \\
\hline
Firefox DNS  & - & - & - & 100 & 98.6 & \textbf{99.3} & 100 & 91.3 & 95.4 \\
\bottomrule[1.5pt]
\end{tabular}
}
\vspace{-0.1cm}
\end{table*}

\subsection{Explicability of Event Semantic Inference}\label{subsec:Interpretability}
We measure the semantics learned by SmartGuard for audit events in both visual and quantitative ways: Visually, we use t-SNE to embed sampled behavior vectors into a 2D plane, giving us an intuitive understanding of the embedding distribution, and we construct five example behaviors to specifically demonstrate the similarity between individual behaviors; Quantitatively, for real audit events, we use normal behavior abstraction (log text summary + embedding vector) and behavior abstraction using only text summary, respectively, to fine-tune the model and compare the detection results.
\

\begin{figure}[t]
    \centering
    \begin{minipage}[t]{0.48\linewidth}
        \centering
        \includegraphics[width=\linewidth]{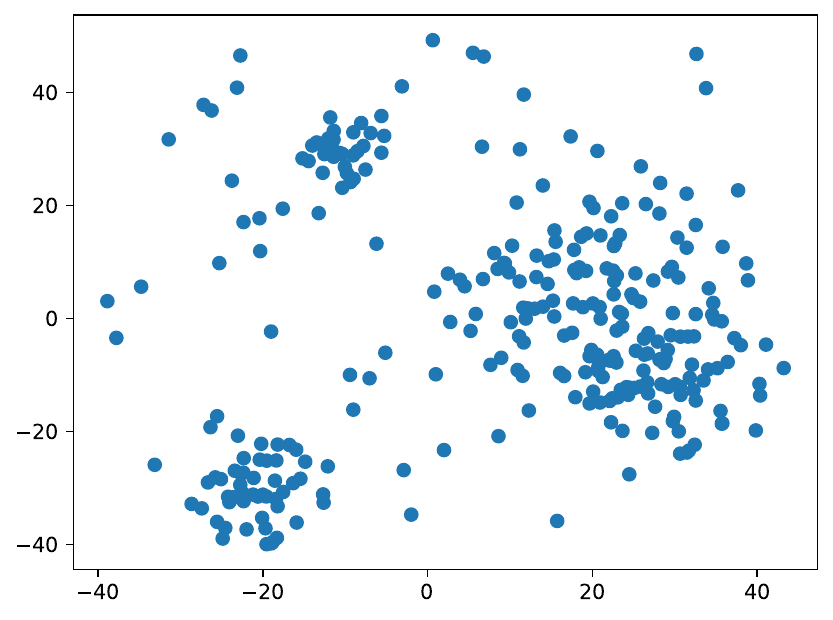}
        \caption*{(a)}
        \label{fig:subfig-a}
    \end{minipage}
    \hfill
    \begin{minipage}[t]{0.48\linewidth}
        \centering
        \includegraphics[width=\linewidth]{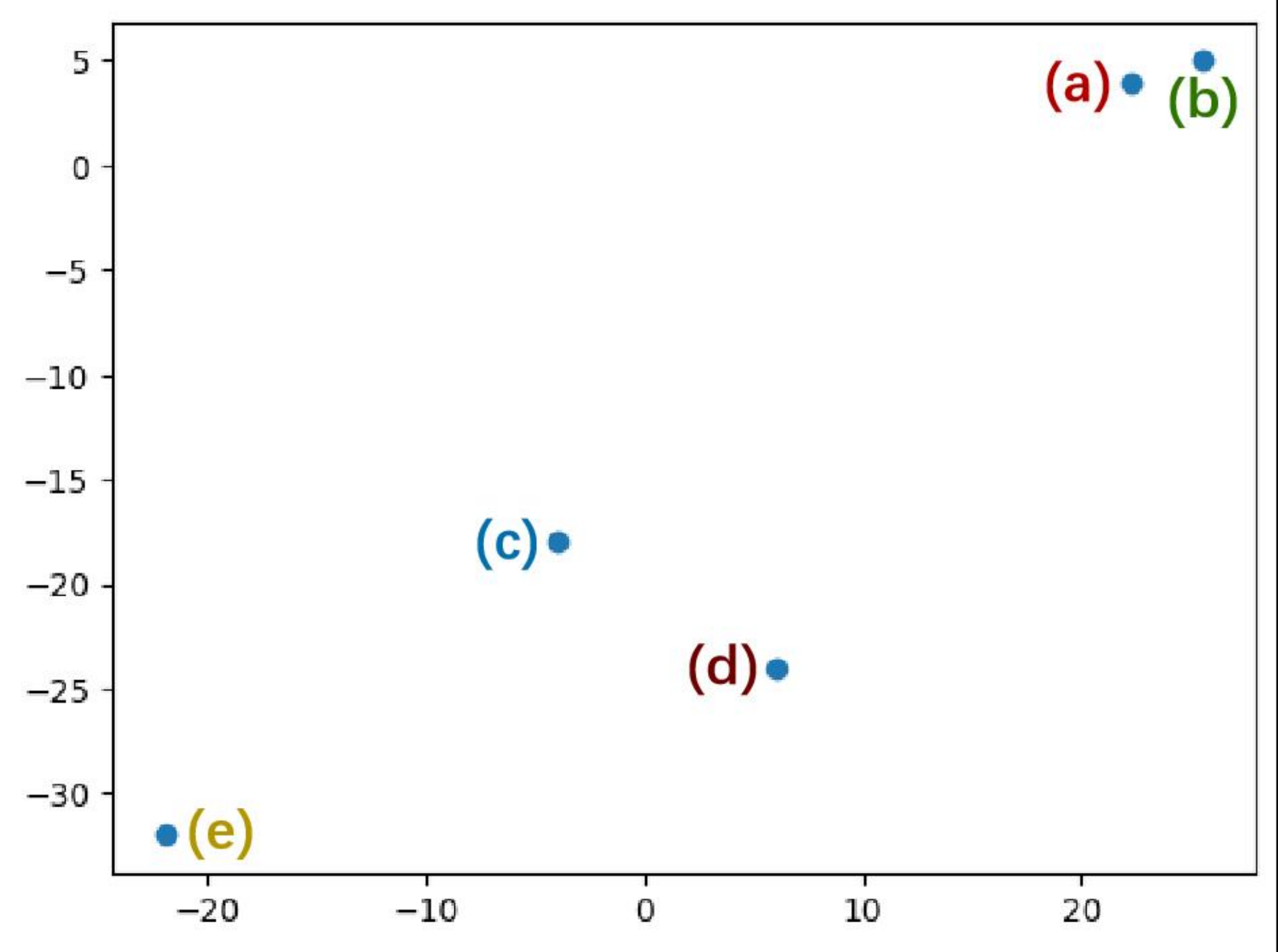}
        \caption*{(b)}
        \label{fig:subfig-b}
    \end{minipage}
    \vfill
    \begin{minipage}[t]{0.98\linewidth}
        \centering
        \includegraphics[width=\linewidth]{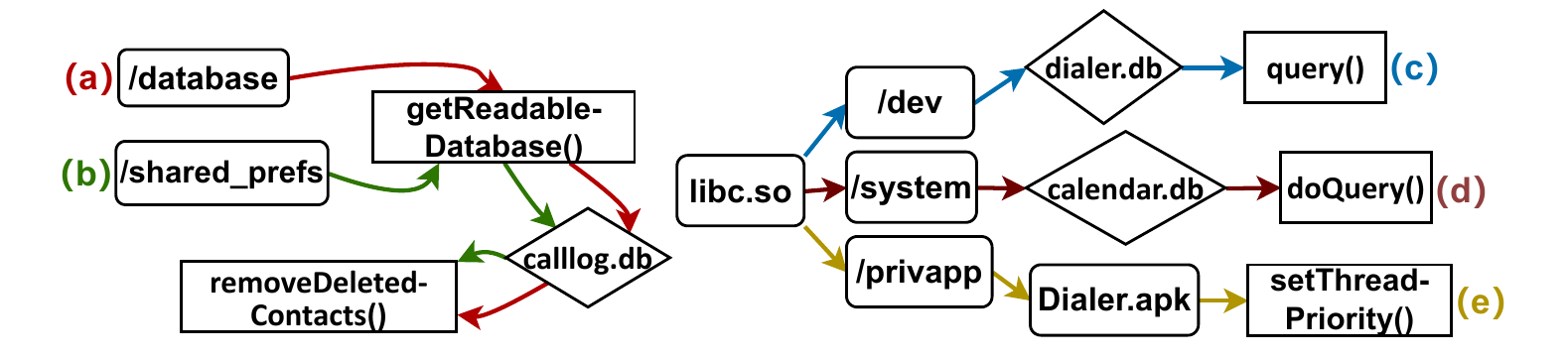}
        \caption*{(c)}
        \label{fig:subfig-c}
    \end{minipage}
    \caption{Visualization of Behavior Abstraction's Embeddings.}
    \label{fig:tsne}
    \vspace{-0.3cm}
\end{figure}

\vspace{0.3em}
\noindent\textbf{Embeddings Visualization.} To evaluate semantic similarity, we projected high-dimensional graph vectors into a 2D space using t-SNE, preserving structural relationships. As shown in Figure \ref{fig:tsne}(a), 300 randomly sampled behaviors from nine scenarios form clustered distributions, indicating meaningful semantic grouping.

We further constructed five behavior instances (Figure \ref{fig:tsne}(c)) to compare specific similarities. Instances a and b—accessing calllog.db from different addresses—exhibit high similarity and belong to the same behavior type. Instances c and d, performing queries on different databases, show moderate similarity. Instance e (launching an APK with high priority) is dissimilar to others.

The scatter plot (Figure \ref{fig:tsne}(b)) confirms that semantically similar behaviors lie closer in the projected space, demonstrating the effectiveness of SmartGuard’s behavior abstraction.


\begin{table}[t]
\tabcolsep=1mm
\renewcommand{\arraystretch}{1.05}
\centering
\caption{Results of the impact of embedding on the accuracy of behavior abstraction.
The \textbf{bold part} indicates the best detection metric F1-score for each category.
}
\label{tab:impact}
\scalebox{0.8}{
\begin{tabular}{c|c|c|c|c|c|c}
\toprule[1.5pt]
\multirow{2}{*}{Category}  & \multicolumn{3}{c|}{SmartGuard w/o embedding(\%)} & \multicolumn{3}{c}{SmartGuard(\%)} \\
\cline{2-7}
 & Precision & Recall  & F1-score  & Precision & Recall & F1-score \\
\hline
Barephone  & 85.2 & 86.9 & 86.1 & 95.9 & 94.5 & \textbf{95.2}  \\
\hline
CADETS Nginx  & 89.1 & 87.2 & 88.2 & 100 & 100 & \textbf{100}  \\
\hline
Metasploit  & 81.5 & 83.2 & 82.4 & 87.6 & 94.9 & \textbf{91.1}  \\
\bottomrule[1.5pt]
\end{tabular}
}
\vspace{-0.3cm}
\end{table}



\vspace{0.3em}
\noindent\textbf{Impact of Embeddings on Behavior Recognition.}
Besides visualization, we also conducted experiments to determine whether the graph embedding vectors affect the behavior abstraction detection results. Specifically, we fine-tuned a large language model in three attack scenarios using normal behavior abstractions and behavior abstractions with only log text summary content. During the inference phase, we used the corresponding behavior abstractions and the large language model for inference, and the results are shown in Table \ref{tab:impact} below. We used the OPT-1.3b model for the experiments.

Table \ref{tab:impact} provides a summary of the experimental results. Compared to detecting only log text summaries, the average F1-score for detecting normal behavior abstractions is 10\% higher. Although text summaries can express some behavioral characteristics of logs, the length and complexity of the summaries limit the large language model's contextual understanding. Therefore, using only text summaries does not yield good inference results. In contrast, graph embedding vectors can express the overall characteristics of behavior with a fixed input length, complementing text summaries and compensating for the inability to express contextual features, thus performing better in behavior diagnosis.

In summary, the behavior abstractions learned by SmartGuard can express the contextual characteristics of events, which is crucial for information diagnosis.

\subsection{Accuracy and Scalability of SmartGuard}\label{subsec:effect}
To evaluate the accuracy of SmartGuard in behavior detection, we use two fine-tuned large language models, Facebook OPT-1.3b, and Open LLaMa2-3b, to predict sessions with similar malicious behavior in the DARPA dataset. For this purpose, we selected the aforementioned 9 behaviors as candidates for behavior abstraction. For each behavior, Table \ref{tab:acc} details the evaluation results of SmartGuard in detecting behavior abstraction in sessions. Performance metrics are measured using precision, recall, and F1 score. Intuitively, they provide measures of true positive rate, false positive rate, and overall accuracy, respectively. We define true positives as sessions where malicious behavior is correctly predicted and false positives as sessions where normal behavior is incorrectly predicted.

\begin{table}[t]
\tabcolsep=2.5mm
\renewcommand{\arraystretch}{1.1}
\centering
\caption{Results of the scalability of SmartGuard on unknown attacks.The results only indicate the accuracy of determining whether it is an \textbf{`Attack Behavior'}.
The \textbf{bold part} indicates the best detection metric F1-score for each category.
}
\label{tab:Scalability}
\scalebox{0.90}{
\begin{tabular}{c|c|c|c|c|c|c}
\toprule[1.5pt]
\multirow{2}{*}{Category}  & \multicolumn{3}{c|}{OPT-1.3b(\%)} & \multicolumn{3}{c}{LLaMa2-3b(\%)} \\
\cline{2-7}
& P & R  & F1  & P & R & F1 \\
\hline
CADETS Nginx  & 91.2 & 85.1 & 88.1 & 92.4 & 87.3 & \textbf{89.8}  \\
\hline
Metasploit & 87.6 & 88.4 & 88.0 & 88.1 & 89.6 & \textbf{88.8}  \\
\hline
Webshell & 87.9 & 86.2 & 87.1 & 87.1 & 87.7 & \textbf{87.4}  \\
\bottomrule[1.5pt]
\end{tabular}
}
\vspace{-0.5cm}
\end{table}

Table \ref{tab:acc} summarizes the experimental results. SmartGuard achieves strong performance in behavioral abstraction detection, with an average F1 score of 95.9\% and consistently outperforms the baseline \cite{satvat2021extractor}. It attains an F1 of 91.6\% even on complex behaviors by leveraging contextual event semantics to accurately infer behavioral intent. For instance, even when an attacker renames system files and uses different functions, SmartGuard correctly associates them with malicious operations based on contextual similarity.

In most cases (14/18), precision exceeds or equals recall, indicating robustness to residual noise after preprocessing. Lower recall in some scenarios (e.g., Metasploit and Webshell) stems from higher noise levels from repeated payload variations and link-scan operations. Nonetheless, SmartGuard maintains a low false positive rate overall.

The method also demonstrates good scalability across two model variants, achieving average F1 scores of 95.6\% and 96.3\%, attributable to high-quality behavior semantics and the reasoning capacity of large language models.

To test the scalability of SmartGuard on the dataset, we retrained the model. We used \texttt{CADETS Nginx} and \texttt{Micro BinFmt-Elevate} from the original dataset as the test set, and the remaining seven attack types as the training set. This was to evaluate whether SmartGuard could classify unseen malicious behaviors as attacks by learning the relationships between behaviors. The specific experimental results are shown in Table \ref{tab:Scalability}. On the scalability of the dataset, SmartGuard showed promising results. The average F1-scores of the two types reached 89.4\%, which is close to the recognition accuracy of the baseline on the full dataset. This indicates that SmartGuard can learn the semantic relationships between different behaviors, demonstrating good scalability on different datasets.

To evaluate SmartGuard’s scalability in detecting unknown attacks, we retrained the model using six attack types for training and tested on three complex types—\texttt{CADETS Nginx}, \texttt{Metasploit}, and \texttt{Webshell}—which performed weakest in the baseline. SmartGuard achieved an average F1 score of 89\%, approaching baseline performance on the full dataset, demonstrating its ability to generalize semantic behavior patterns to unseen attacks.

Furthermore, by fine-tuning with expert-provided traceability graphs from three additional attack types, SmartGuard reached an F1 score of 94.3\% on the original test set, surpassing the baseline and closing within 2–3\% of the full-dataset model. These results confirm that SmartGuard can efficiently incorporate new attack semantics and enable effective detection in real-world deployments.


\begin{table}[t!]
\tabcolsep=2.5mm
\renewcommand{\arraystretch}{1.1}
\centering
\caption{Results of SmartGuard's detection after fine-tuning for unknown attacks using expert-provided traceability graphs. The model was fine-tuned with expert traceability graphs different from the original data, and the detection metric is the attack type.
The \textbf{bold part} indicates the best detection metric F1-score for each category.
}
\label{tab:Scalability2}
\scalebox{0.90}{
\begin{tabular}{c|c|c|c|c|c|c}
\toprule[1.5pt]
\multirow{2}{*}{Category}  & \multicolumn{3}{c|}{OPT-1.3b(\%)} & \multicolumn{3}{c}{LLaMa2-3b(\%)} \\
\cline{2-7}
& P & R  & F1  & P & R & F1 \\
\hline
CADETS Nginx  & 98.4 & 97.5 & 97.9 & 98.9 & 97.4 & \textbf{98.1}  \\
\hline
Metasploit & 87.9 & 93.6 & 90.7 & 90.2 & 93.7 & \textbf{91.9}  \\
\hline
Webshell & 92.9 & 92.9 & 92.9 & 95.5 & 97.1 & \textbf{96.3}  \\
\bottomrule[1.5pt]
\end{tabular}
}
\vspace{-0.3cm}
\end{table}




In summary, SmartGuard can identify malicious behaviors and distinguish them from benign ones. It can also learn the semantic relationships between different behaviors, achieving good recognition accuracy for unknown attacks not present in the training set. Analysts can further enhance SmartGuard's ability to recognize specific behaviors by fine-tuning with instructions and expert-provided traceability graphs for unknown malicious behaviors. This offers a new approach to solving real-world problems.

\subsection{Interpretability of LLMs and Solutions for Hallucinations}\label{subsec:cot}

We employ Chain-of-Thought (CoT) prompting to enable large language models to interpret audit events, using a hierarchical instruction structure to improve comprehension of behavioral summaries. This approach guides the model to identify sensitive nodes, critical links, and overall behavioral traits, providing sufficient contextual logic for accurate judgment and explanatory narratives. Finally, the model classifies the attack type and offers detailed explanations.

We take the behavior graph in Figure \ref{fig:example1} as an example to illustrate the CoT process shown in Figure \ref{app:cot}. As can be seen from the figure, the large language model can make accurate and interpretable narratives of the behavior summary, thereby further reducing the workload of analysts.

Furthermore, CoT can mitigate the issue of hallucinations in large models to some extent. We tested single-turn response (the original mode of SmartGuard, which directly judges the attack type) and multi-turn response (first judging sensitive nodes and links, then determining if it is an attack behavior, and finally identifying the specific attack type) on the OPT-1.3b model during the behavior detection phase. Table \ref{tab:cot} shows the experimental results. The results indicate that the CoT-based multi-turn response can effectively improve the accuracy of SmartGuard and reduce false positives caused by hallucinations. We observed a significant increase in the precision metric compared to single-turn response, mainly due to the reduction in false positive samples, i.e., fewer samples were mistakenly classified as attack behaviors. 
\begin{table}[t]
\tabcolsep=2.5mm
\renewcommand{\arraystretch}{1.15}
\centering
\caption{Results of mitigating hallucinations of SmartGuard. 
The \textbf{bold part} indicates the best detection metric F1-score for each category.
}
\label{tab:cot}
\scalebox{0.90}{
\begin{tabular}{c|c|c|c|c|c|c}
\toprule[1.5pt]
\multirow{2}{*}{Category}  & \multicolumn{3}{c|}{Single-turn(\%)} & \multicolumn{3}{c}{Multi-turn(\%)} \\
\cline{2-7}
& P & R  & F1  & P & R & F1 \\
\hline
JAVA APK  & 94.7 & 93.9 & 94.3 & 97.6 & 93.9 & \textbf{95.7}  \\
\hline
Metasploit  & 87.6 & 94.9 & 91.1 & 93.2 & 95.8 & \textbf{94.4}  \\
\hline
Firefox Drakon  & 100 & 95.6 & 97.8 & 100 & 96.1 & \textbf{98.0}  \\
\hline
Webshell  & 92.4 & 93.3 & 92.8 & 94.8 & 94.2 & \textbf{94.5}  \\
\bottomrule[1.5pt]
\end{tabular}
}
\vspace{-0.3cm}
\end{table}

In summary, our results show that SmartGuard can provide a certain level of explainability and mitigate the misjudgments caused by hallucinations to some extent.

\section{Discussion}\label{sec:dis}

\vspace{0.3em}


\vspace{0.3em}
\noindent\textbf{RQ1: How is the methodology of SmartGuard different from previous methods?}

In addition to the use of LLM, compared with previous process-level behavior extraction solutions, SmartGuard reduces the granularity of extraction to the thread level and adds function-level behavioral semantic records, not just system scheduling. In the traditional audit log analysis process, triples (Subject, Relation, Object) are recorded at the process level. This method can effectively and concisely extract a high-level behavior, but it will ignore the more detailed parts behind the behavior and improve the difficulty of detection. 
Then we added LLM for pattern recognition, and aggregated text summaries and graph embedding vectors into description information so that LLM can perform reasoning and large language models can more smoothly understand behavior summaries and make accurate judgments while also being able to give explanatory narratives of specific steps in a behavior. Additionally, SmartGuard has advantages in detecting unknown attacks and supports timely system updates.

\vspace{0.3em}
\noindent\textbf{RQ2: How is the robustness of Behavior Abstraction?}

To evade behavior abstraction, attackers may attempt to confuse behavior by deliberately introducing irrelevant events. However, the impact of such events on behavior semantics is limited. In Sections \ref{subsec:3.3.2} and \ref{subsec:3.3.3}, we introduce two strategies (thread-based behavior partitioning and noise events) to enhance SmartGuard's robustness to abstract behavior. Specifically, while SmartGuard aggregates the contextual semantics of behavior, irrelevant events will be assigned low-importance weights or not be partitioned into specific behaviors, and may even be deleted as noise events. Another potential method to counteract behavior obfuscation is to incorporate additional auxiliary information (e.g., semantically rich parameters of audit events) into SmartGuard's KG. We believe this can empower SmartGuard with more capabilities to filter out uninteresting events for security analysis.

\vspace{0.3em}
\noindent\textbf{RQ3: Why is the graph embedding space and updating?}

Most learning methods usually use embedding techniques~\cite{ordentlich2016network}, and due to semantic changes and the inclusion of previously unseen data (e.g., data format changes), SmartGuard needs to periodically retrain the embedding space. Additionally, compared to natural language embedding models, graph embedding models have a natural advantage in focusing attention on key nodes and restoring the entire event scenario. That is, the graph embedding method we chose can fully express behavior semantics for pattern recognition.

\section{Related Works}\label{sec:related works}

\noindent \textbf{Causal Analysis. }Causal analysis, though orthogonal to behavior abstraction, provides critical support by enabling reasoning over log events. King and Chen \cite{king2005enriching} pioneered the use of dependency graphs for tracking security events to their root causes. Subsequent work by Jin et al. \cite{kwon2018mci} improved causal tracking with forward and cross-host dependencies. To address challenges like dependency explosion and high storage costs, researchers have proposed techniques including fine-grained unit partitioning \cite{lee2013high, ma2015accurate, ma2017mpi}, model-based reasoning \cite{kwon2016ldx, kwon2018mci}, record and replay \cite{ji2017rain, ji2018enabling}, and general provenance \cite{hassan2020omegalog}. Another line of work reduces log volume via graph compression \cite{chen2017distributed, hassan2018towards, hossain2018dependence, tang2018nodemerge} and data reduction \cite{lee2013loggc, ma2016protracer, xu2016high}. While SmartGuard operates in a different scope, it relies on accurate causal analysis for correlating events and summarizing behavior.


\vspace{0.3em}
\noindent \textbf{Behavior Abstraction. }Abstracting system behaviors into graphs or causal dependencies has proven effective for understanding OS-level activities and detecting threats. Prior work includes TGMiner \cite{zong2015behavior}, which mines discriminative graph patterns as templates; POIROT \cite{milajerdi2019poirot}, which aligns APT behavior graphs from threat reports; SLEUTH \cite{hossain2017sleuth} and MORSE \cite{hossain2020combating}, which model information leakage via labeling strategies; and HOLMES \cite{milajerdi2019holmes} and RapSheet \cite{hassan2020tactical}, which represent multi-stage attacks as TTP-aligned causal chains. In contrast, SmartGuard abstracts behavior through context-based embeddings and log text integration, enabling quantitative semantic representation and advanced behavioral analysis.


\vspace{0.3em}
\noindent \textbf{Embedded Space and Large Language Model Analysis. }Embedding techniques have been widely applied to log analysis tasks such as anomaly detection \cite{chen2016entity, liu2019log2vec, manzoor2016fast}, malware identification \cite{wang2020you, wang2019heterogeneous}, and modeling attack evolution \cite{shen2019attack2vec}. Common approaches employ neural networks, word embeddings, or n-grams to convert logs into vector representations. For instance, DeepLog \cite{du2017deeplog} uses LSTM to model normal log sequences, while PROVDETECTOR \cite{wang2020you} applies doc2vec to quantify process behaviors. ATTACK2VEC \cite{shen2019attack2vec} leverages temporal word embeddings to represent attack steps, and Extractor \cite{satvat2021extractor} constructs provenance graphs using semantic and syntactic features. UNICORN \cite{han2020unicorn} introduces graph sketching for summarizing system executions, and WATSON \cite{zeng2021watson} uses TransE to capture contextual semantics.
Unlike these methods, SmartGuard combines graph vectors and natural language summaries to represent behavioral context, integrating structural and textual semantics. In contrast to RCACopilot \cite{chen2024automatic}, which relies on GPT-4 and large log corpora, SmartGuard can operate effectively on smaller models without pre-trained log data, offering greater flexibility and lower deployment cost for analysts.

\section{Conclusion}\label{sec:conclusion}
Abstracting high-level behaviors from low-level audit logs is a critical task in security response. It helps bridge the semantic gap between audit events and system behaviors, thereby reducing the manual effort in log analysis. In this paper, we propose an automated method, SmartGuard, to extract behaviors from audit events and analyze them using large language models. Specifically, SmartGuard utilizes context information from log-based knowledge graphs for semantic inference. To differentiate representative behaviors, SmartGuard provides a representation method that combines behavior semantics graph vectors with text summaries. It uses this to perform reasoning with a large language model and provides interpretive narratives. We evaluated SmartGuard against adversarial engagement behaviors organized by DARPA. The experimental results show that SmartGuard can accurately abstract benign and malicious behaviors, demonstrating good scalability for unknown attacks. Additionally, the provided interpretive narratives and timely system updates effectively address real-world scenarios.




\bibliographystyle{ACM-Reference-Format}
\bibliography{references}

\appendix

\section*{appendix}\label{sec:appendix}

\section{Implementation Details}\label{app:implementationDetail}
SmartGuard is implemented in Python 3.8 with around 2.5K lines of code (LoC).
 In this section, we discuss important technical details in the implementation.

\vspace{0.3em}
\noindent \textbf{Audit Log Input Interface.}
SmartGuard takes system audit data as input. We defined a general interface for audit logs and built input drivers to support different log formats, such as the Linux Audit \cite{linux_audit_kernel} format (cdm20 and DARPA dataset).

\vspace{0.3em}
\noindent \textbf{Knowledge Graph Construction.}\label{subsubsec:pt}
To construct a log-based Knowledge Graph (KG), SmartGuard first sorts audit events in chronological order. Then, it divides the records of different threads according to the thread ID. It then converts each event into a triplet using system entities as the Head and Tail, and system call functions as the Relation. To facilitate the comparison of different nodes, we classify each type of element, such as file\_path and function in system entities, and execute, link, check\_attribute in system calls. Therefore, in the graph neural network, different weights can be assigned according to the categories of nodes and edges. After parsing the audit events into triplets, the NEO4J \cite{neo4j} tool is used to store and construct the KG.

\vspace{0.3em}
\noindent \textbf{Embedding Generation.}
We first use the \textit{word2vec} model and the \textit{Gensim} library \cite{rehurek2010software} to generate node embeddings for training the graph. The length of the initial node embedding vectors is limited to 100 to reduce the model size. A two-layer \textit{GAT} model is used to aggregate graph information. The input and output dimensions of the node vectors in the model are 100 and 128, respectively. We optimize the model parameters using the \textit{Adam} optimizer. We train the model for 20 epochs with a batch size of 64. The learning rate changes exponentially with a rate of 0.98, starting at 0.01.

\vspace{0.3em}
\noindent \textbf{Computational Hardware.}\label{subsubsec:ft}
We use Pytorch \cite{paszke2019pytorch} as the backend. All experiments are performed on an Ubuntu 20.04 system equipped with a 96-core Intel CPU and four Nvidia GeForce RTX A6000 GPUs.

\vspace{0.3em}
\noindent \textbf{Dataset.}\label{subsec:dataset}
The DARPA TC dataset \cite{transparent_computing_e5} is a publicly available APT attack dataset released under the DARPA Transparent Computing (TC) program \cite{darpa_transparent_computing}. The dataset originates from the host network during a two-week red team vs. blue team engagement 3 in April 2018. During the engagement, enterprises simulated various security-critical services such as web servers, SSH servers, email servers, and SMB servers \cite{milajerdi2019holmes}. The red team conducted a series of nation-state and common attacks on target hosts while performing benign activities such as SSH logins, web browsing, and email checking. The DARPA TRACE dataset consists of 726,072,596 audit events, forming 211 graphs. Overall, we use nine of these attack scenarios to evaluate the interpretability and accuracy of SmartGuard. We introduce the scenarios and attributes in Table \ref{tab:scenario1}. The first column represents the name of the attack scenario, the second column provides a brief description of the scenario, and the third and fourth columns represent the average number of edges and nodes in the corresponding behavior graph.

For each attack scenario, we construct corresponding subgraphs, including benign and malicious source graphs. Each type of traceability graph (including benign traceability graphs) has a sample size of 4000, with 90\% used as the training set.

\section{Training Details of GAT Model}\label{app:lossfunction}
The loss function for optimizing the embedding model is:
\[
\mathcal{L} = -\sum_{(i,j) \in \mathbf{KG}} \log \sigma(\mathbf{h}_i , \mathbf{h}_j) - \sum_{(i,j) \notin \mathbf{KG}} \log (1 - \sigma(\mathbf{h}_i , \mathbf{h}_j)) ],\tag{3}
\]
where \( \mathbf{KG} \) is the set of edges in the graph. \( \mathbf{h}_i \) and \( \mathbf{h}_j \) are the embedding vectors of nodes \( i \) and \( j \), respectively. \( \sigma(\cdot) \) represents the cosine similarity function, which maps the cosine similarity to the range [0, 1] for easier computation.

Additionally, we also optimize the attention pooling model. This is a single-layer Fully Connected layer (FC) that obtains the vector representation of the entire graph by performing a weighted sum of the node embedding vectors. We use the same loss function as in Equation (2) and optimize it together with the GAT model during training.

\section{CoT Process}\label{app:cot}
LLMs have demonstrated exceptional ability in understanding the context of downstream tasks and generating relevant information from demonstrations. Therefore, we use the CoT (Chain of Thought) approach to enable large language models to interpret audit events. We designed a hierarchical instruction structure to help the large language model better understand the behavior summary, make accurate judgments, and provide explanatory narratives of specific steps.
Specifically, we let the large language model judge and summarize the sensitive nodes, sensitive links, and overall characteristics of the behavior. This is because such CoT instructions provide the large language model with sufficient contextual logic, enabling it to answer most of the behavior detail questions needed by analysts and make accurate judgments. Finally, we let the large language model determine the attack category and provide more explanations for this attack type.

\begin{figure}[t]
  \centering
   \begin{tcolorbox}[colback=gray!30, colframe=gray!30, coltitle=black, arc=4pt, boxrule=0.5pt, boxsep=2pt, left=2pt, right=2pt, top=2pt, bottom=2pt, before skip=0.5\baselineskip, after skip=0.5\baselineskip]
 \textbf{Instruction:} The following description shows the log information of the event. It contains a summary of the log text and behavior Embeddings.  Please determine whether there are any sensitive nodes in this event. If so, please specify them.
 
\textbf{LLM Output:} libmicroapt.so,barephoneinstr.apk,mmssms.db,
removeDeletedContacts(),insertUpdateContacts().\\
 \\
  \textbf{Instruction:} The following description contains a summary of the log text and behavior Embeddings. Please indicate the \textbf{sensitive behavior} path based on the sensitive nodes. The answer template is `From...to...'.
 
\textbf{LLM Output:} From libmicroapt.so to insertUpdateContacts().\\
\\
  \textbf{Instruction:} The following description contains a summary of the log text and behavior Embeddings. Please indicate the \textbf{overall characteristics} of the sensitive behavior.
 
\textbf{LLM Output:} Load libmicroapt, install barephoneapk, and perform multiple deletions and updates on the mmssms.db database.\\
\\
  \textbf{Instruction:} The following description contains a summary of the log text and behavior Embeddings. Please focus on the Embeddings to determine \textbf{whether this event is a malicious event}. If it is, please provide the classification of the malicious event. 
 
\textbf{LLM Output:} Yes, the category is Barephone Micro.\\
\\
  \textbf{Instruction:} The following description contains a summary of the log text and behavior Embeddings. Please provide \textbf{more explanations} about this attack behavior. 
 
\textbf{LLM Output:} Barephone Micro Attack generally loads Micro APT by installing an apk, and the target of the attack is usually to steal database information.

 \end{tcolorbox} 
  \caption{Interpretive narrative of the behavior summary by the large language model.}
  \label{fig:cot}
\end{figure}

\section{Graph Construction Example}\label{app:graph}
\begin{figure}[t]
    \centering
    \includegraphics[width=0.49\textwidth]{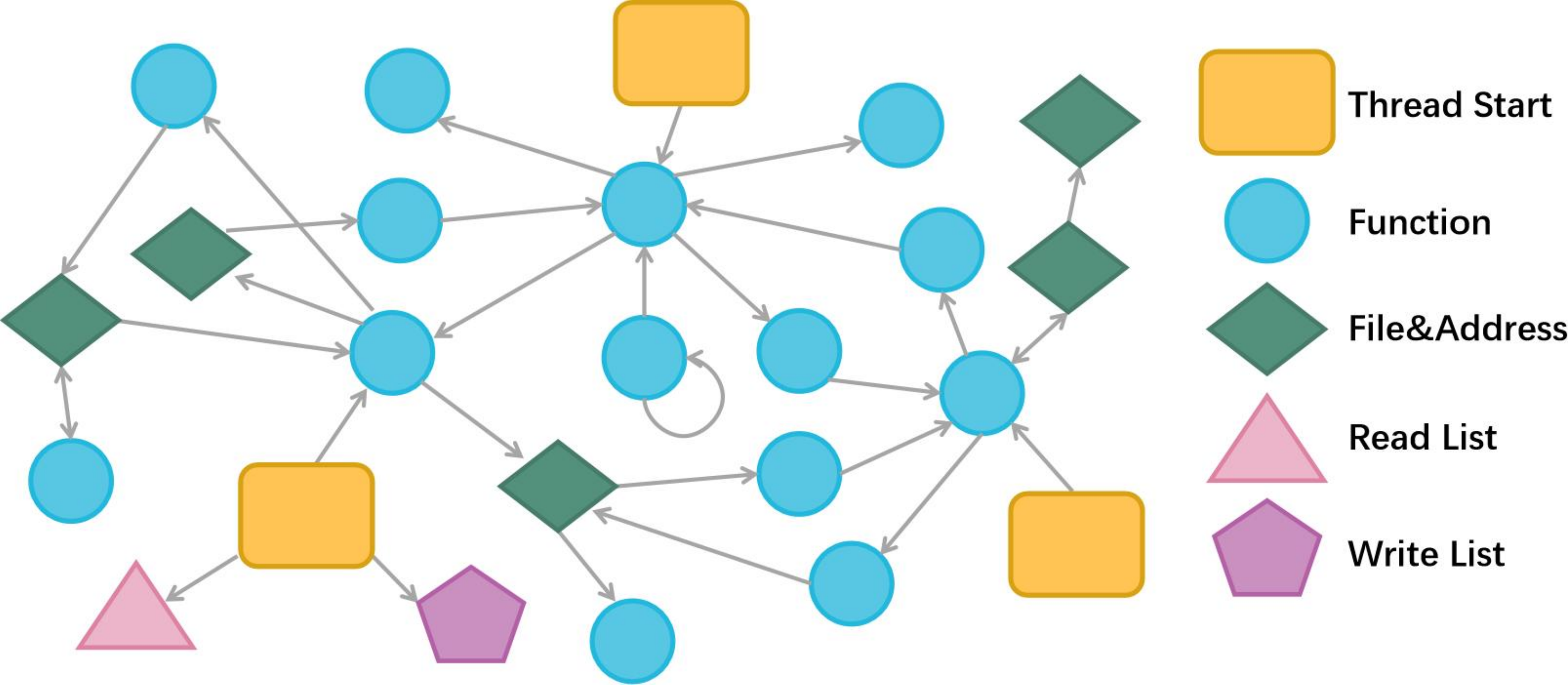} 
    \caption{Example of Graph Construction. We use different colors and shapes to mark different nodes, corresponding to the node types in Table \ref{tab:node_example}. The lines between the nodes represent system calls.}
    \label{fig:graph}
\end{figure}

Following the formal description of KG by Färber et al. \cite{farber2018linked}, we define the log-based KG as a Resource Description Framework (RDF) graph \cite{miller1998introduction}. More specifically, the log-based knowledge graph is a set of numerous semantic triples. Each triple corresponds to an audit event and consists of three elements, encoding the semantic relationship between \textit{Head} and \textit{Tail} in the form of (\textit{Head}, \textit{Relation}, \textit{Tail}). Both \textit{Head} and \textit{Tail} can be any type of system entity, mainly including specific file addresses, functions, read/write lists, etc. \textit{Relation} can be any system operation performed on \textit{Tail}, such as READ, LINK, and MODIFY\_FILE\_ATTRIBUTES. Table \ref{tab:node_example} presents some examples of our classification of node types. Additionally, we divide the logs by threads, with each thread subgraph starting with a thread ID and connecting under the process ID that created them. Threads are interconnected through the same entities, such as the same addresses and functions. Figure \ref{fig:graph} illustrates an example of one of the attack behavior subgraphs.

\section{Node Types}\label{app:node}
Table \ref{tab:node_example} presents some examples of our classification of node types. It includes five types of nodes, among which \texttt{Function} and \texttt{File or Address} are the most significant for determining behavior. The \texttt{Readlist} and \texttt{Writelist}, while containing read/write operations, typically only record benign activities such as login records.A \texttt{Thread Start} node denotes the beginning of a thread and specifies the thread's ID. This node is linked to the \texttt{Readlist} and \texttt{Writelist}. The count of \texttt{Thread Start} nodes in an event represents the number of threads whose behaviors are incorporated into that event.
\begin{table}[t]
\renewcommand{\arraystretch}{1.15}
\centering
\caption{Classification of node types and some specific examples of nodes.
}
\label{tab:node_example}
\scalebox{1.0}{
\begin{tabular}{c|c}
\toprule[1.5pt]
Node Type & Node Name \\
\hline
Thread Start & thread id \\
\hline
Function & query(),setThreadPriority(),execTransact()  \\
\hline
File or Address  & libc.so,dialar.db,/system,/dev/ashmem  \\
\hline
Readlist  &  printlnnative \\
\hline
Writelist  & writeEvent \\
\bottomrule[1.5pt]
\end{tabular}
}
\end{table}

\begin{table*}[t!]
\tabcolsep=4mm
\renewcommand{\arraystretch}{1.1}
\centering
\caption{Overview of attack cases in DARPA TC dataset with scenario descriptions}
\label{tab:scenario1}
\scalebox{0.95}{
\begin{tabular}{>{\centering\arraybackslash}p{2cm} | p{0.7\textwidth} | >{\centering\arraybackslash}p{0.4cm} | >{\centering\arraybackslash}p{0.4cm}}
\toprule[1.5pt]
\textbf{Attack Case} & \multicolumn{1}{|c|}{\textbf{Scenario Description}} & \textit{V$|$G$|$} & \textit{E$|$G$|$}\\
\hline
APK Java &  The Java APT automatically connects to the target network and runs privilege escalation software to gain root access and steal files. & 25 & 36\\
\hline
Barephone Micro & An APT that runs on a mobile phone typically connects to the target network by loading \texttt{libmicroapt.so}. & 10 & 16\\
\hline
CADETS Nginx & Nginx vulnerability was exploited to attack CADETS
FreeBSD. The attacker downloaded a file, elevated it
to a new process running as root, and attempted to
inject it into the \texttt{sshd} process. & 17 & 28\\
\hline
 Firefox Drakon & The attacker exploits a Firefox vulnerability to attack CADETS through malformed HTTP requests, downloads the \texttt{libdrakon} implant file \texttt{*.so}, and injects it into the \texttt{sshd} process, causing CADETS to crash.
 & 26 & 31\\
\hline
Metasploit APK & The attacker uses Metasploit as malware to send a malicious executable file to the target host and attack ClearScope. & 35 & 56\\
\hline
Micro BinFmt-Elevate & The attacker uses the \texttt{ta1-pivot-2} tool to achieve BinFmt elevation and gain access to the root directory. & 14 & 34\\
\hline
AppStarter APK & Typically disguised as a legitimate AppStarter APK file, it can use benign activities to install privilege escalation programs. & 12 & 19\\
\hline
Webshell & The shell connection is made to the operator console via HTTP. The attacker executed the apt implant without root privileges. & 31 & 49\\
\hline
Firefox DNS FileFilter & The attack triggered a Firefox backdoor to connect to the target network via DNS. By searching for processes that open specific non-existent files, the attacker elevated privileges.
 & 23 & 29 \\
\bottomrule[1.5pt]
\end{tabular}
}
\vspace{-0.3cm}
\end{table*}

\section{Attack cases in DARPA TC dataset with scenario descriptions}\label{app:attackCase}
We use nine of these attack scenarios to evaluate the interpretability and accuracy of SmartGuard. We introduce the scenarios and attributes in Table \ref{tab:scenario1}. It includes the name/type of the attack behavior, a description of its characteristics, and the average values of nodes and edges used to construct the graph.


\section{Ethics Statement}
Our solution is a semantic extraction and inference detection scheme based on audit logs, which is a defense scheme against cyber attacks. In this paper, we do not introduce new threats or vulnerabilities. Our experimental data is all sourced from the open-source DARPA dataset, and no personally identifiable information (PII), such as IP addresses, was collected. As such, this work does not raise any major ethical issues.


\end{document}